\documentclass[aps,prb,twocolumn,groupedaddress,showpacs]{revtex4}
\begin{document}
\title{Diamagnetic response of Aharonov-Bohm rings: impurity backward scatterings}
\author{Mun Dae Kim,$^1$ Chul Koo Kim,$^2$ and Kyun Nahm$^3$}
\affiliation{$^1$ Korea Institute for Advanced Study, Seoul 130-722, Korea\\
$^2$ Institute of Physics and Applied Physics, Yonsei University, Seoul 120-749, Korea\\
$^3$ Department of Physics, Yonsei University, Wonju 220-710, Korea}
\date{\today}

\begin{abstract}
We report a theoretical calculation on the persistent currents of
disordered normal-metal rings. It is shown that the
diamagnetic responses of the rings in the vicinity of the zero
magnetic field are attributed to multiple backward scatterings off the
impurities. We observe the transition from the
paramagnetic response to the diamagnetic one as the strength of disorder
grows using both the analytic calculation and the
numerical exact diagonalization.
\end{abstract}

\draft
\pacs{}
\maketitle

\section{Introduction}

The phase coherence of electrons in  mesoscopic rings gives rise to
many interesting quantum phenomena. Among them the persistent
current in clean rings induced by the Aharonov-Bohm (AB) flux has been extensively studied
and is now well understood both theoretically and experimentally.
However, recently, new puzzling behaviors have been observed for the averaged persistent
currents in diffusive rings. Experiments clearly indicate that the averaged persistent
currents show diamagnetism \cite{Levy,Jariwala,DeblockL,DeblockB} in the  vicinity of the zero magnetic field
in contrary to the existing theory
and the amplitude of the persistent current \cite{Chandras} is much larger than
the theoretically predicted values. \cite{Cheung,Oppen}

There have been several attempts to explain these puzzling behaviors.
The repulsive electron-electron interactions may enhance the amplitude
of the persistent current, but produce only paramagnetic responses for the ensemble averaged currents.
\cite{AmbegPRL,Groeling} The nonequilibrium ac noise
in a mesoscopic ring can cause both the decoherence of electron wave
function and the diamagnetic dc current in the loop. \cite{Yudson}
Including the spin-orbit scattering in a diffusive ring with
ac noise, it was  claimed
that the sign of persistent current changes from diamagnetic to
paramagnetic as the strength of the spin-orbit scattering
increases.\cite{Kravtsov} However experiments show that the persistent currents
of  Au rings \cite{Jariwala} and Ag rings \cite{DeblockL} with
strong spin-orbit interactions as well as of GaAs rings
\cite{DeblockB} with a weak one all exhibit the diamagnetic
responses. Even though the experiments were performed at much
higher temperature than the superconducting transition
temperatures, it was claimed that the superconducting fluctuations
due to the phonon mediated attractive interaction might lead to a
diamagnetic magnetization. \cite{AmbegEPL} Furthermore an attempt
which included the large contribution of far levels from the Fermi level
\cite{Schechter} claimed that superconducting  fluctuations would
give a much larger persistent current.

The magnetic response of an AB ring depends on the parity of
the number of electrons $n$ in the ring. For $n=4N$ the states of the topmost electrons,
the clockwise moving and the counterclockwise moving states, are degenerate when
AB flux $\Phi_{\rm ext}$ is zero.  The paramagnetic current emerges when the degeneracy
becomes lifted due to a finite $\Phi_{\rm ext}$ and, thus, the circulating direction of
the topmost electrons is determined.
The ensemble average of the currents of rings with $n=4N, n=4N \pm 2$ and $n=4N \pm 1$
show paramagnetic responses with the periodicity $\Phi_0/2$ in clean limit
or in weakly disordered case, where $\Phi_0=h/e$ is the unit flux quantum.
However, in a disordered ring, the backward scattering of  electrons off the impurities
may induce transitions between the two states and thus severely reduce the paramagnetic
persistent current for $n=4N$.

In the disordered systems the backward scattering process occurs through multiple scattering
process\cite{Mello} which leads to rich interesting physical phenomena
such as weak localization \cite{Bergmann} and universal conductance fluctuation. \cite{Lee}
Experiments for diffusive normal-metal rings \cite{ring} as well as wires \cite{wire} showed that
the resistance oscillation along with the external magnetic field can be explained in terms of the weak
localization theory. \cite{Alt} Therefore we can think that there is much higher
backscattering probability in those normal-metal rings.
A numerical study showed that the amplitude of backscattering in  a disordered multi-channel wire
increases along with the length of the wire and the reflection coefficient  grows up to
become comparable to the transmission coefficient. \cite{Mochales}

In present study we show that the backscattering process in the diffusive ring
induce the transition between the topmost levels and, as a result,
the ensemble  averaged persistent current changes from paramagnetic to diamagnetic
as the impurity strength  grows.

\section{diamagnetism due to backscattering process}


The Hamiltonian of an AB ring with potential impurity scattering can be given by
\begin{eqnarray}
H=\sum_{k_m} \epsilon_{k_m} c^\dagger_{k_m} c_{k_m}
+ \sum_{k_m,p} V_p c^\dagger_{k_m+p} c_{k_m},
\end{eqnarray}
where $k_m=2\pi(m+f)/L$ and $p=2\pi m'/L$ with $f\equiv \Phi_{\rm
ext}/\Phi_0$, the circumference of the ring $L$ and integers $m$
and $m'$. Here  $V_p$ is the Fourier component of the impurity
potential $v(x)$. The scattering term of the Hamiltonian,
$\sum_{k_m,p} V_p c^\dagger_{k_m+p} c_{k_m}$, can be divided into
the forward scattering and the backscattering process depending on
the value of the momentum transfer $p$.
For a diffusive AB ring the forward scattering  may reduce the amplitude of
persistent current but cannot transfer  electrons to oppositely moving states.
Thus, although we will consider all the scattering processes
in the following exact diagonalization calculation,
we first consider only the backward scattering process
for transparency of argument.

\begin{figure}[b]
\vspace*{5cm}
\includegraphics{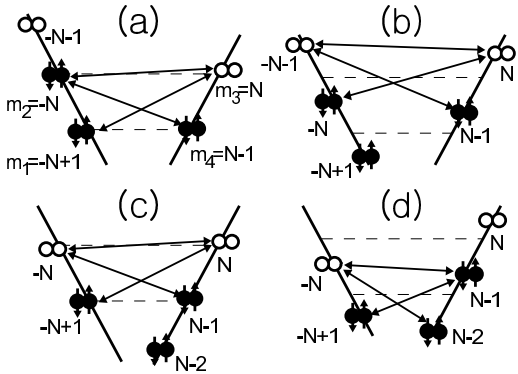}
\vspace*{0.5cm}
\caption{Energy levels
of the AB ring. When $n=4N$, (a) shows the levels for $f\approx 0$ and
(b) for $f\approx 0.5$. (c) and (d) correspond to the case $n=4N-2$.
Here filled (open) circles denote the occupied (unoccupied) levels.}
\label{scatt}
\end{figure}

The backscattering process in an AB ring  can be described by
$c^\dagger_{-k_m+q} c_{k_m}$, where $q=2\pi(n+2f)/L$ with integer $n$.
When there is no threading AB flux and thus $q=2\pi n/L$,
it is sufficient to consider the most dominant process such as $c^\dagger_{-k_m} c_{k_m}$
for $q=0$ with $n=0$ at ground state.
However, since the value of $q$ cannot be zero for finite AB flux due to the symmetry breaking
of the momentum levels, we need to consider the backscattering process with finite $q$.

For  a multichannel wire the backscattering processes  take place through multiple scatterings
between the states in different channels, which accompany momentum transfer between the initial
and the final state. When an electron becomes backscattered eventually,
the effective backscattering process can be given by $w(q) c^\dagger_{-k_m+q}c_{k_m}$.
If there is no AB flux threading the loop, the effective backscattering can be described as
$\omega(q)c^\dagger_{-k_m} c_{k_m}$ approximately with $\omega(q) =\omega_0 \delta_{q,0}$.
However, for finite $\Phi_{\rm ext}$, $\omega(q)$ will have some distribution which is centered
about  $q \approx 0$.  For example, the backscattering processes between edge states in quantum Hall
system through multiple scatterings off impurities show a Lorentzian distribution \cite{Gurvitz}
and light scatterings from a one-dimensional rough random metal a Gaussian-type distribution
for the direction of the reflected waves. \cite{Leskova}

Here we take  a Gaussian-type distribution for $\omega(q)$ such as
\begin{eqnarray}
\label{distr}
\omega(q)=\omega_0 e^{-\alpha q^2a^2},
\end{eqnarray}
where $a$ is the lattice constant of the loop
and  we set $\alpha=0.1$ here and after.
Different values of $\alpha$ or different distribution functions
result in only quantitatively different behaviors.
%
Then the effective  single channel Hamiltonian where the
backscattering term comes from multiple scatterings between
different channels can be written as follows,
\begin{eqnarray}
H=\sum_{k_m} \epsilon_m c^\dagger_{k_m} c_{k_m} + \sum_{k_m,q} \omega(q) c^\dagger_{-k_m+q} c_{k_m},
\end{eqnarray}
where
$\epsilon_m \equiv E_0(m+f)^2$ and $E_0\equiv
2\pi^2\hbar^2/m_eL^2$ with the mass of an electron $m_e$.
%

Since the scatterings off the impurities occur mainly for the electrons near the Fermi level,
it is sufficient to consider the scattering processes between several levels
near the Fermi level as shown in Fig. \ref{scatt}.
The contribution of other electrons to  the persistent current
are obtained assuming that they do not participate in the scattering processes.

When an electron is backscattered from the state, $|k_m\rangle$, to the
state, $|k_{m'}\rangle$, the scattering matrix element becomes
$W_{mm'}\equiv  \omega(q)$ with $k_{m'}=-k_m+q$.
The Green's function $G_{mm}(\epsilon)$ can be obtained from the equation
$(\epsilon \pm is-H)G=I$ which can be written as follows,
\begin{eqnarray}
\label{Green}
(\epsilon\pm is -\epsilon_m)G_{mm'}(\epsilon) ~~~~~~~~~~~~~~~~~~~~~~~~~~~~~~~~~~\nonumber\\
-\sum_{m''\neq m}W_{mm''}G_{m''m'}(\epsilon)=\delta_{mm'}.~~~~~
\end{eqnarray}
In Fig. \ref{scatt}(a),   we consider scatterings between four
levels, $m_1=-N+1, m_2=-N, m_3=N$, and $m_4=N-1$,
with a small external flux, $f \approx 0$.
Including the electrons at the zero momentum state, we can see that the total number of
electrons $n$ in Fig. \ref{scatt}(a) is $4N$.
The scattering processes are expressed as arrows in Fig. \ref{scatt}(a).

The Green's functions can be obtained from Eq. (\ref{Green}) as
follows,
\begin{eqnarray}
G_{m_2m_2}^{-1}(\epsilon)=\epsilon \pm is -\epsilon_{m_2} ~~~~~~~~~~~~~~~~~~~~~~~~~~~~~~~~~~~~\nonumber\\
\label{G22}
-\frac{G_{m_4m_2}(\epsilon)}{G_{m_2m_2}(\epsilon)}W_{m_2m_4}
-\frac{G_{m_3m_2}(\epsilon)}{G_{m_2m_2}(\epsilon)}W_{m_2m_3},~~~~~~~~~~~\\
G_{m_4m_4}^{-1}(\epsilon)= \epsilon \pm is -\epsilon_{m_4}~~~~~~~~~~~~~~~~~~~~~~~~~~~~~~~~~~~~\nonumber\\
\label{G44}
-\frac{W^2_{m_2m_4}}{\epsilon \pm is -\epsilon_{m_2}
-\frac{G_{m_3m_4}(\epsilon)}{G_{m_2m_4}(\epsilon)}W_{m_2m_3}},~~~~~~~~~~~~~~~~
\end{eqnarray}
where
\begin{eqnarray}
\frac{G_{m_2m_2}(\epsilon)}{G_{m_2m_4}(\epsilon)}=\frac{1}{W_{m_2m_4}}(\epsilon \pm is -\epsilon_{m_4}),
~~~~~~~~~~~~~~~~~~~~\\
\frac{G_{m_2m_4}(\epsilon)}{G_{m_3m_4}(\epsilon)}=\frac{G_{m_2m_2}(\epsilon)}{G_{m_2m_3}(\epsilon)}
~~~~~~~~~~~~~~~~~~~~~~~~~~~~~~~~~~~~\nonumber\\
=\frac1{W_{m_2m_3}}\left[\epsilon \pm is -\epsilon_{m_3}-\frac{W^2_{m_1m_3}}{(\epsilon \pm is -\epsilon_{m_1})}\right].
~~~~~
\end{eqnarray}
The Green's functions $G_{m_3m_3}(\epsilon)$ and $G_{m_1m_1}(\epsilon)$
can be obtained from Eqs. (\ref{G22}) and (\ref{G44})
by exchanging the momentum indices; $m_2\leftrightarrow m_3$ and $m_4\leftrightarrow m_1$.

All these four Green's functions have the same denominator ${\mathcal D}$ represented as
\begin{eqnarray}
{\mathcal D}= {\mathcal F_{m_1m_3}}{\mathcal F_{m_2m_4}} ~~~~~~~~~~~~~~~~~~~~~~~~~~~~~~~~~~~~~~\nonumber\\
- (\epsilon \pm is -\epsilon_{m_1})(\epsilon \pm is -\epsilon_{m_4})W^2_{m_2m_3} ~~~
\end{eqnarray}
with
\begin{eqnarray}
{\mathcal F_{m_im_j}} \equiv (\epsilon \pm is -\epsilon_{m_i}) (\epsilon \pm is -\epsilon_{m_j})-W^2_{m_im_j},
\end{eqnarray}
from which we can obtain four poles $\epsilon^k$ such that
$\epsilon^1\leq\epsilon^2\leq\epsilon^3\leq\epsilon^4$.
The persistent currents are calculated using the residues of the Green's functions
obtained by $z^k_{m_i}=\lim_{\epsilon \rightarrow \epsilon^k}(\epsilon-\epsilon^k)G_{m_im_i}(\epsilon)$.
Since we consider three scattering particles in four states of Fig. \ref{scatt}(a), the probability
that a certain level, $m_i$, is occupied at the ground state is represented as the sum of the residues of the three lowest
energy states, $p_{m_i}=\sum^{n_p}_{k=1} z^k_{m_i}$ with $n_p=3$.

Since electrons which are not shown in the figure are assumed not to
participate in the scattering process, the contributions
of these electrons can be obtained  simply summing
the momentum of each electron;\cite{Loss}
\begin{eqnarray}
I'=-\frac{e}{L}\sum_m \frac{2\pi}{L}(m+f)=-\frac{I_0}{N}\sum_m (m+f)
\end{eqnarray}
with $I_0\equiv ev_F/L$ and $v_F \equiv 2\pi N/L$.
Therefore, the total persistent current is written as the sum of the two contributions as follows,
\begin{eqnarray}
\label{Itot}
I=-2\frac{I_0}N \left[\sum^4_{i=1} p_{m_i}(m_i+f)+\sum^{m_r}_{m=m_l}(m+f)\right],
\end{eqnarray}
where the first part represents the persistent current of the three scattered electrons
in Fig. \ref{scatt}(a) and the second one that of the unscattered electrons from $m_l=-N+2$ to $m_r=N-2$.
The prefactor 2 comes from the spin degeneracy.

When the AB flux $f$ increases to  $f \approx 0.5$, the energy
levels become shifted as shown in Fig. \ref{scatt}(b). In this
case, since the energy levels of the two states, $|N\rangle$ and
$|-N-1\rangle$, are much closer than before, the scattering
process between these two states becomes important unlike to the case of Fig. \ref{scatt}(a).
For the persistent  current of Fig. \ref{scatt}(b),
we use the definition of parameters; $m_1=-N+1, m_2=-N, m_3=N$, $m_4=N-1$, $m_l=-N+1$,
$m_r=N-2$ and $n_p=2$. Here and after we set $N=100$.

\begin{figure}[t]
\vspace*{6cm}
\includegraphics{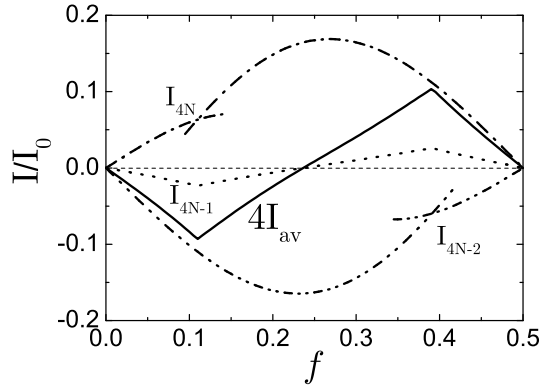}
\vspace*{0cm}
\caption{Persistent currents for $\alpha=0.1$ and $\omega_0/\Delta=0.7$.
Left (right) curves in $I_{4N}$ and $I_{4N-2}$ correspond to Fig. \ref{scatt}(a) and
Fig. \ref{scatt}(c) (Fig. \ref{scatt}(b) and Fig. \ref{scatt}(d)), respectively.
We show $4I_{av}$ for the ensemble averaged persistent current instead of
$I_{av}$ for clarity.}
\label{GreenPc}
\end{figure}

Since we consider a quadratic dispersion, $\epsilon_m=E_0(m+f)^2$,
the level spacing  at the Fermi level is of the order of $NE_0$.
Here we define the level spacing $\Delta\equiv NE_0$.
In Fig. \ref{GreenPc} we show the persistent current $I_{4N}$ for
$n=4N$ with $\omega_0/\Delta=\omega_0/NE_0=0.7$ and $\alpha=0.1$,
where the left (right) curve corresponds to Fig. \ref{scatt}(a) (Fig. \ref{scatt}(b)).
In the intermediate regime of the external flux $f$,
we extend the results from the both limits to
obtain the approximate results as shown in Fig. \ref{GreenPc}.

%
When the total number of electrons is $n=4N-2$, the persistent current $I_{4N-2}$  can also be
obtained similarly using the scattering processes shown in Fig. \ref{scatt}(c) and (d),
where we use the parameters $m_1=-N+1, m_2=-N, m_3=N$, $m_4=N-1$, $m_l=-N+2, m_r=N-2$
and $n_p=2$ for Fig. \ref{scatt}(c) and  $m_1=-N+1, m_2=-N, m_3=N-1$, $m_4=N-2$,
$m_l=-N+2, m_r=N-3$ and $n_p=3$ for Fig. \ref{scatt}(d).

The dotted line in Fig. \ref{GreenPc} shows the persistent current $I_{4N-1}$ for
$n=4N-1$ which is obtained by the relation $I_{4N-1}=0.5(I_{4N}+I_{4N-2})$.
This relation can be easily understood for clean loops without impurities.
If we consider one additional electron at $m=-N$ state
in Fig. \ref{scatt}(c), say spin-up electron, it becomes the configuration
for $n=4N-1$ with the number of spin-up electrons $n_u=2N$ and
spin-down electrons $n_d=2N-1$. The configuration for the spin-up electrons
is the same as that of Fig. \ref{scatt}(a) with half the number of electrons
and, for the spin-down electrons, it corresponds to Fig. \ref{scatt}(c).
Since we do not have spin flip processes, the total persistent
current is equal to the sum of  half the  persistent currents of $n=4N$ and $n=4N-2$.
In the present case of loops with impurities, if the impurity scattering is not
spin dependent so that each spin degree of freedom is decoupled from the other,
we have the same relation.

The current $I_{4N+1}$ for $n=4N+1$ is the same as that of $I_{4N-1}$ with
small difference of $O(1/N)$. Hence, the ensemble averaged persistent
current $I_{av}$ can be written as
\begin{eqnarray}
I_{av}=\frac14 (I_{4N}+I_{4N-2}+2I_{4N-1}).
\end{eqnarray}
The ensemble averaged persistent current  in clean rings  shows the paramagnetic
responses for $f\approx 0$, since the paramagnetic currents of $I_{4N}$
dominate over the diamagnetic current $I_{4N-2}$.
Even for an AB loop with impurities, the qualitative picture does not change
although the scattering is expected to reduce the magnitude of the paramagnetic
current,  if one considers only the forward scatterings off the impurities.
However, when we consider contributions from the backward scattering processes,
the situation changes drastically as indicated in Fig. \ref{GreenPc}.

It is well known that the paramagnetic response  for $n=4N$ comes from the
contribution of the electrons with $m_2=-N$ in Fig. \ref{scatt}(a).
At $f=0$  the levels at $m_2=-N$ and $m_3=N$ in Fig. \ref{scatt}(a) are degenerate.
Finite external flux lifts the degeneracy and the
electrons at $m_2=-N$ contribute a large magnitude of persistent current in addition to
the currents by the other electrons in the lower levels, thus resulting in
the paramagnetic ensemble averaged current.
However, when the backward impurity scattering is sufficiently strong,
the impurity scattering causes mixing of the two levels. As a result,
the population probability of the level $m_2=-N$ decreases, whereas that
of the level $m_3=N$ increases. This reduces the paramagnetic $I_{4N}$
and results in the diamagnetic averaged current $I_{av}$  for small $f$ in Fig. \ref{GreenPc}.

\begin{figure}[b]
\vspace*{6.5cm}
\includegraphics{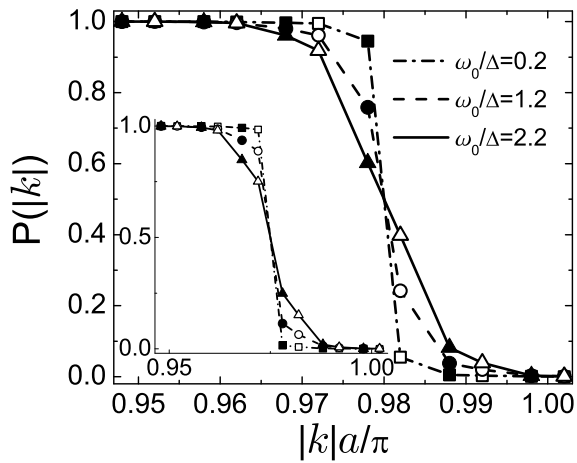}
\vspace*{-0.cm}
\caption{Momentum distribution of the AB loop for $n=4N$ with 12 levels of which
7 levels are filled when $f=0.1$. The horizontal axis denote the absolute value of the wave vector
in the loop. The filled (open) marks show the occupation probability of the levels with negative
(positive) values of wave vectors. Inset shows the case for $n=4N-2$ with 12 levels of which
6 levels are filled.}
\label{MD}
\end{figure}

\section{numerical calculation}

Since we considered impurity scatterings of only four levels in calculating the Green's functions,
the persistent current curves in Fig. \ref{GreenPc} are only approximate.
In order to obtain more accurate persistent current curves we perform
a numerical calculation using the exact diagonalization method,
where we include the forward scatterings, $\sum_{k_m,q} w(q) c^\dagger_{k_m+q} c_{k_m}$, as well as the backscatterings.
We consider a many-body basis with twelve levels near the Fermi level,
where seven (six) levels are filled for $n=4N ~(n=4N-2)$.
If there is no impurity, the ground state is $|111111100000\rangle$ ($|111111000000\rangle$)
in the occupation-number space for $n=4N$ $(n=4N-2)$.
Here we does not consider spin degree of freedom of electrons and the amplitude of the persistent
current  will be multiplied by the spin degeneracy at the end of calculation.

We calculate the ground state by diagonalizing the Hamiltonian  matrix with 792 (924) basis for several values of $\omega_0$
for $n=4N$ $(n=4N-2)$ and  show the calculated momentum distribution  for $n=4N$
as a function of the absolute value of momentum level when $f=0.1$ in Fig. \ref{MD}.
The filled marks denote the levels with negative momentum (left branch) and
the open marks with positive momentum (right branch).
For weak impurity strength, we note that the lowest seven levels are almost occupied.
The seven levels are composed of four levels in the left branch and three levels in the right branch.
In this case, it is easily seen that the contribution from the
topmost level with $ka/\pi=-0.978$ in left branch becomes dominant.

\begin{figure}[t]
\vspace*{9.5cm}
\includegraphics{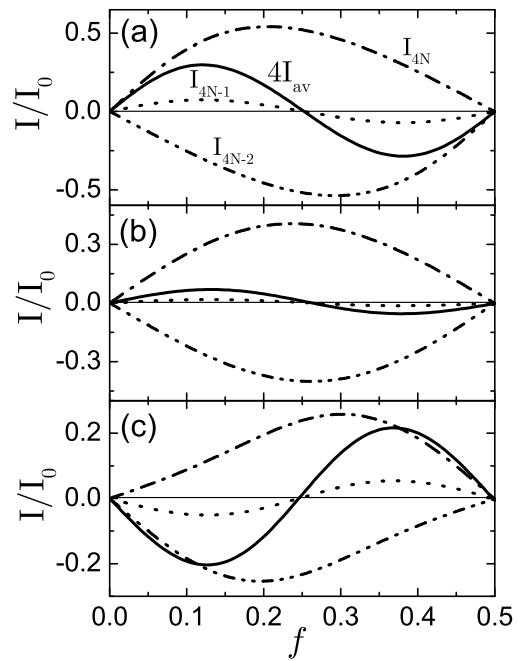}
\vspace*{0cm}
\caption{Persistent currents for (a) $\omega_0/\Delta=1.2$, (b) $\omega_0/\Delta=1.6$ and
(c) $\omega_0/\Delta=2.0$ obtained by the exact diagonalization method with 12 levels.
$I_{4N-1}$ (the dotted line) is obtained using the relation $I_{4N-1}=0.5(I_{4N}+I_{4N-2})$.}
\label{a01pc}
\end{figure}

In Fig. \ref{a01pc}(a), we show a paramagnetic persistent current for weak impurity
scattering strength $\omega_0/\Delta=1.2$, where the paramagnetic persistent current $I_{4N}$ dominates over the
diamagnetic one $I_{4N-2}$. Here, the contributions of the electrons which do not participate
in the scattering processes to the persistent
currents are included as in the second term of Eq. (\ref{Itot}).
However, as the strength of the impurity increases, the occupation probability of the topmost level
in the left branch decreases while that of the topmost level
with $ka/\pi=0.982$ in the right branch increases as can be seen in Fig. \ref{MD}.
The persistent currents due to these two states cancel each other and only the difference
of the population probability $P(|k|=0.978)-P(|k|=0.982)$ contributes to the paramagnetic response.
In Fig. \ref{MD}, this corresponds to the difference between the filled mark
and the open mark for each value of $\omega_0$ about $ka/\pi \approx0.98$.

Figure \ref{a01pc}(b) shows that the persistent current changes from the
paramagnetic  to the diamagnetic one as the strength of the impurities increases.
When $\omega_0/\Delta=2.0$ in Fig. \ref{a01pc}(c), the difference $P(|k|=0.978)-P(|k|=0.982)$ for $n=4N$
becomes much smaller, whereas it is still very large for $n=4N-2$ as shown in the inset of
Fig. \ref{MD}. Therefore, the persistent current $I_{4N}$ becomes  weak  compared to
that of $I_{4N-2}$ and, thus, the ensemble averaged persistent current shows the diamagnetic response
as observed in the experiments. \cite{Levy,Jariwala,DeblockL,DeblockB}
A numerical simulation similar to the previous study \cite{Mochales} may be possible.
Using the twisted boundary condition for  the wave functions in an AB ring with a disordered zone inserted
we think that it can confirm the present results.


\section{summary}

In summary, we have presented a theoretical calculation on the sign of ensemble averaged
persistent current for the disordered AB rings.
The experimentally observed diamagnetic response is attributed to the transition
process of the topmost electrons to the opposite branch for $n=4N$.
It is shown that the backward scatterings of the topmost electrons
give rise to the diamagnetic currents
in a natural way in agreement with the recent experimental observations.


\begin{center}
{\bf ACKNOWLEDGMENTS}
\end{center}

This work was supported in part by Korea Research Foundation Grant No.
KRF-2003-005-C00011.

\end{document}